\documentclass{PoS}

\title{Scattering length from BS wave function inside the interaction range}

\ShortTitle{Scattering length from BS wave function inside the interaction range}

\author{\speaker{Y.Namekawa}
        \\
        Faculty of Pure and Applied Sciences,
        University of Tsukuba, Tsukuba, Ibaraki 305-8571, Japan
        \\
        E-mail: \email{namekawa@het.ph.tsukuba.ac.jp}
       }

\author{T.Yamazaki
        \\
        Faculty of Pure and Applied Sciences,
        University of Tsukuba, Tsukuba, Ibaraki 305-8571, Japan
        \\
        Center for Computational Sciences, University of Tsukuba,
        Tsukuba, Ibaraki 305-8577, Japan
        \\
        E-mail: \email{yamazaki@het.ph.tsukuba.ac.jp}
       }

\abstract{
 We evaluate $I=2$ two-pion scattering length through the scattering amplitude obtained by the Bethe-Salpeter wave function inside the interaction range.
 The scattering length is computed with $m_\pi = 0.52-0.86$ GeV in the quenched lattice QCD.
 Furthermore, the half-off-shell amplitude is calculated, from which the effective range is extracted.
 Our results are compared with those by the conventional finite size method and by chiral perturbation theory to confirm consistency.
}

\FullConference{The 36th Annual International Symposium on Lattice Field Theory - LATTICE2018\\
		22-28 July, 2018\\
		Michigan State University, East Lansing, Michigan, USA.}

\begin{document}

\section{Introduction}

Lattice simulation is a substantial technique for an understanding of hadrons quantitatively from the first principle of QCD.
Lattice QCD provides scattering information, such as the scattering length $a_0$, the effective range $r_{\rm eff}$, and the scattering phase shift $\delta(k)$ itself.
The recent lattice QCD results are reviewed in Ref.~\cite{Briceno:2018bnl}.

A conventional lattice QCD approach to the scattering information utilizes the finite volume formula and its extensions,
originally proposed by M.L\"{u}scher~\cite{Luscher:1986pf}.
The formula describes a relation between two-hadron energy in a finite box and $\delta(k)$ in the infinite volume by an analytic function.
In quantum field theory, this formula is derived using the Bethe-Salpeter (BS) wave function outside the two-hadron interaction range $R$~\cite{Lin:2001ek,Aoki:2005uf}.
A method to obtain a potential from Schr\"{o}dinger-type equation by the BS wave function is also suggested~\cite{Aoki:2009ji}.

An associated issue is raised in the infinite volume,
which connects the BS wave function inside $R$ and the on-shell scattering amplitude~\cite{Lin:2001ek,Aoki:2005uf,Yamazaki:2017gjl}.
This issue motivates us to perform a lattice QCD simulation for scattering amplitudes from the BS wave function inside $R$~\cite{Namekawa:2017sxs}.
The simulation employed the isospin $I=2$ S-wave two-pion system at the pion mass $m_\pi = 0.86$~GeV in the quenched QCD.
Our result of the on-shell amplitude confirmed consistency between the finite volume method and our approach.
We also obtained the half-off-shell scattering amplitude successfully by lattice QCD.
It is not observable in experiments, but it provides additional input for hadron effective theories and models, complementary to the experimental data.
Furthermore, we can extract the effective range from the half-off-shell scattering amplitude under two assumptions.

In this proceedings, we investigate the quark mass dependence of the scattering amplitudes.
Chiral extrapolations are performed with data ranged in $m_\pi = 0.52-0.86$~GeV at the lattice spacing of $a^{-1} = 1.207$~GeV.
We confirm consistency between the previous result by the conventional finite size formula and our result at each simulation point and at the physical point.
Comparison with the phenomenological value is also exhibited.

\section{Formulation on the lattice}

We employ the notation in our previous paper~\cite{Namekawa:2017sxs}, following Refs.~\cite{Lin:2001ek,Aoki:2005uf,Yamazaki:2017gjl}.
The BS wave function of $I=2$ two pions on the lattice $\phi({\bf x};k)$ is extracted from the two-pion four-point function,
\begin{equation}
 \langle 0 | \Phi({\bf x},t) | \pi^+ \pi^+, E_k \rangle 
 = C_k \phi({\bf x};k) e^{-E_k t} + \cdots,
 \label{eqn:four_point_func}
\end{equation}
where $\cdots$ are excited state contributions.
$C_k$ is an overall constant.
$E_k = 2 \sqrt{m_\pi^2 + k^2}$ is the two-pion energy.
$| \pi^+ \pi^+, E_k \rangle$ is a ground state of two pions.
$\Phi({\bf x},t)$ is a two-pion operator defined by $\Phi({\bf x},t) = \sum_{{\bf r}} \pi^+(R_{A_1^+}[{\bf x}] + {\bf r}, t) \pi^+({\bf r}, t)$,
using a single pion interpolating operator, $\pi^{+}({\bf x},t) = \bar{d}({\bf x},t) \gamma_5 u({\bf x},t)$.
$A_1^+$ projection $R_{A_1^+}[{\bf x}]$ is applied to obtain the S-wave scattering at the center of mass.
Contribution from higher angular momentum $l \geq 4$ is also involved, but negligible.

Once $\phi({\bf x};k)$ is calculated, the reduced BS wave function $h({\bf x};k)$ can be determined by
\begin{equation}
 h({\bf x};k) = (\Delta + k^2) \phi({\bf x};k),
 \label{eqn:def_hxk_lat}
\end{equation}
where the symmetric lattice laplacian is used, $\Delta f({\bf x}) = \sum_{i=1}^{3} ( f({\bf x}+\hat{i}) + f({\bf x} - \hat{i}) - 2 f({\bf x}) )$.
It can be also formulated in the momentum space, as is often employed in the continuum theory~\cite{Carbonell:2016ekx}.
We numerically confirmed both approaches give a consistent result.

The key property of $h({\bf x};k)$ is that outside the interaction range of the two-pion $R$,
\begin{equation}
 h({\bf x};k) = 0 \mbox{\ \ \ for } x > R,
 \label{eqn:sufficient_condition}
\end{equation}
where the exponential tail is supposed to be tiny.
If $R$ is less than a half of the lattice extent $L$ and the exponential tail is negligible,
the half-off-shell amplitude on the lattice $H_L(p;k)$ can be obtained through $h({\bf x};k)$,
\begin{equation}
 H_L(p;k) = -\sum_{{\bf x} \in L^3} C_k h({\bf x};k) j_0(px),
 \label{eqn:H_L_pk_lattice}
\end{equation}
where $j_0(px)$ is the spherical Bessel function.
$R < L / 2$ is the sufficient condition for lattice QCD calculation of the scattering amplitude.
The half-off-shell amplitude in the infinite volume is related to $H(p;k)$ by $H(p;k) = H_L(p;k) / C_{00}$,
where $C_{00} = C_k / F(k,L)$ with a finite volume factor of two pions $F(k,L)$~\cite{Lellouch:2000pv}.
It associates the two-pion BS wavefunction in the infinite volume $\phi_\infty({\bf x};k)$ with $\phi({\bf x};k)$ on the lattice, $\phi_\infty({\bf x};k) = F(k,L) \phi({\bf x};k)$.

The scattering phase shift $\delta(k)$ is related to $H(k;k)$ through
\begin{eqnarray}
 H(k;k) = \frac{4 \pi}{k} e^{i \delta(k)} \sin \delta(k).
 \label{eqn:H_kk_infinite_volume}
\end{eqnarray}
Though the phase factor $e^{i \delta(k)}$ can not be evaluated on the lattice, we can utilize some ratio to obtain $\delta(k)$.
$H_L(k;k) / (C_k \phi({\bf x}_{\rm ref};k) )$ at a reference point ${\bf x}_{\rm ref}$, for example, provides $\delta(k)$,
\begin{equation}
 \tan \delta(k)
 = \frac{\sin(k x_{\rm ref})}
        {4 \pi x_{\rm ref} C_k \phi({\bf x}_{\rm ref};k) / H_L(k;k)
         - \cos(k x_{\rm ref})}.
 \label{eqn:tan_delta}
\end{equation}
The overall constants and the phase factor are canceled in the ratio.
The scattering length $a_0$ and the effective range $r_{\rm eff}$ are defined by the effective range expansion,
\begin{equation}
 \frac{k}{\tan \delta(k)} = \frac{1}{a_0} + r_{\rm eff} k^2 + O(k^4).
 \label{eqn:ERE}
\end{equation}
We estimate $a_0$ approximately by
\begin{equation}
 a_0 = \frac{\tan \delta(k)}{k}.
 \label{eqn:a_0}
\end{equation}
A tiny value of $k^2$ of $I=2$ two-pion justifies this estimation.
Similarly, $r_{\rm eff}$ is estimated by
\begin{eqnarray}
 r_{\rm eff}
 = - \frac{2 k^2 H' + \sin^2 \delta(k)}
            {2 k \sin \delta(k) \cos \delta(k)},
 \ \ \
 H'
 = \frac{\partial H(p;k)}{\partial p^2}|_{p^2 = k^2} / H(k;k),
 \label{eqn:r_eff}
\end{eqnarray}
where we assume $\partial (H(p;k) e^{-i \delta(k)}) / \partial p^2 \sim \partial (H(p;p) e^{-i \delta(p)}) / \partial p^2$ and the phase of $H(p;k)$ is $e^{i \delta(k)}$ at $p^2 \sim k^2$.

\section{Set up}

We perform a quenched QCD simulation as a test bed of our approach for the scattering amplitude.
Gauge configurations are generated on $24^3 \times 96$ and $24^3 \times 64$ lattices, saved at each 100 HMC trajectories.
We employ Iwasaki gauge action~\cite{Iwasaki:2011jk} at the bare coupling $\beta = 2.334$, which corresponds to a lattice spacing of $a^{-1} = 1.207$~GeV~\cite{AliKhan:2001tx}.
We use Clover quark action with a mean field improvement, $C_{\rm SW} = 1.398$.
The valence quark hopping parameters are $\kappa_{\rm val} = 0.1340, 0.1358, 0.1369$.
These hopping parameters corresponds to pion masses of $m_\pi = 0.86$--$0.52$~GeV.
Table~\ref{tab:simulation} compiles our simulation parameters.

The two-pion four-point function is computed with random $Z2$ sources to prevent from Fierz rearrangement.
The source is located at a time slice $t_{\rm src}$ and all spatial points.
The source spreads also in all colors and spins to reduce the simulation cost~\cite{Boyle:2008rh}.
The gain is found to be a factor of three in our simulation.
We use four random sources to calculate all six combinations of a pair of quark propagators from different random sources.
Wall sources are also employed at $\kappa_{\rm val} = 0.1340$ on $24^3 \times 64$ for confirmation of operator independence of our simulation results.
The wall sources are placed at $t_{\rm src}$ and $t_{\rm src} + 1$ to deflect Fierz contamination~\cite{Kuramashi:1993ka}.
The pion two-point functions are calculated in the same set up as those of the two-pion four-point function.
The boundary conditions are chosen to be periodic in spatial directions, and Dirichlet in the temporal direction.
The Dirichlet boundary is placed away from $t_{\rm src}$ by 12-time distance.

Our data is fitted in the range of $[t_{\rm min},t_{\rm max}] = [14,44]$ to determine a single pion mass and two-pion energy on $24^3 \times 64$ lattice,
and $[t_{\rm min},t_{\rm max}] = [14,74]$ on $24^3 \times 96$ lattice.
We employ $[t_{\rm min},t_{\rm max}] = [44,74]$ for a fit of wave functions on $24^3 \times 96$.
On $24^3 \times 64$ lattice, only a time slice of $t = 44$ is available for the wave function analysis.

The interaction momentum $k$ is determined by two methods.
One is the momentum of the two-pion energy of the temporal correlator in Eq.~(\ref{eqn:four_point_func}), denoted by $k_t = \sqrt{ E_k / 4 - m_\pi^2 }$.
The other is the momentum defined by the spatial wave function outside the interaction range, denoted by $k_s = \sqrt{ - \Delta \phi({\bf x};k) / \phi({\bf x};k) }, x > R$.
The sufficient condition of Eq.~(\ref{eqn:sufficient_condition}) is satisfied by definition, $(\Delta + k_s^2) \phi({\bf x};k) = 0$ for $x > R$.
$k_s$ has been found to be more precise than $k_t$~\cite{Aoki:2005uf}.

\begin{table}[tb]
\begin{center}
\begin{tabular}{ccccc}
 \hline \hline
 Lattice size     &  $\kappa_{\rm val}$    &  $N_{\rm src}$  &  Source type  &  $N_{\rm config}$
 \\ \hline
 $24^3 \times 96$ &  0.1340,0.1358,0.1369  &  24             &  Z2           &  200
 \\ \hline
 $24^3 \times 64$ &  0.1340                &  32             &  Z2           &  400
 \\ \hline
                  &                        &  16             &  Wall         &  200
 \\ \hline \hline
\end{tabular}
 \caption{
  \label{tab:simulation}
  Simulation parameters.
 }
\end{center}
\end{table}

\section{Result}

\begin{figure*}[t]
 \centering
 \includegraphics[scale=0.50]{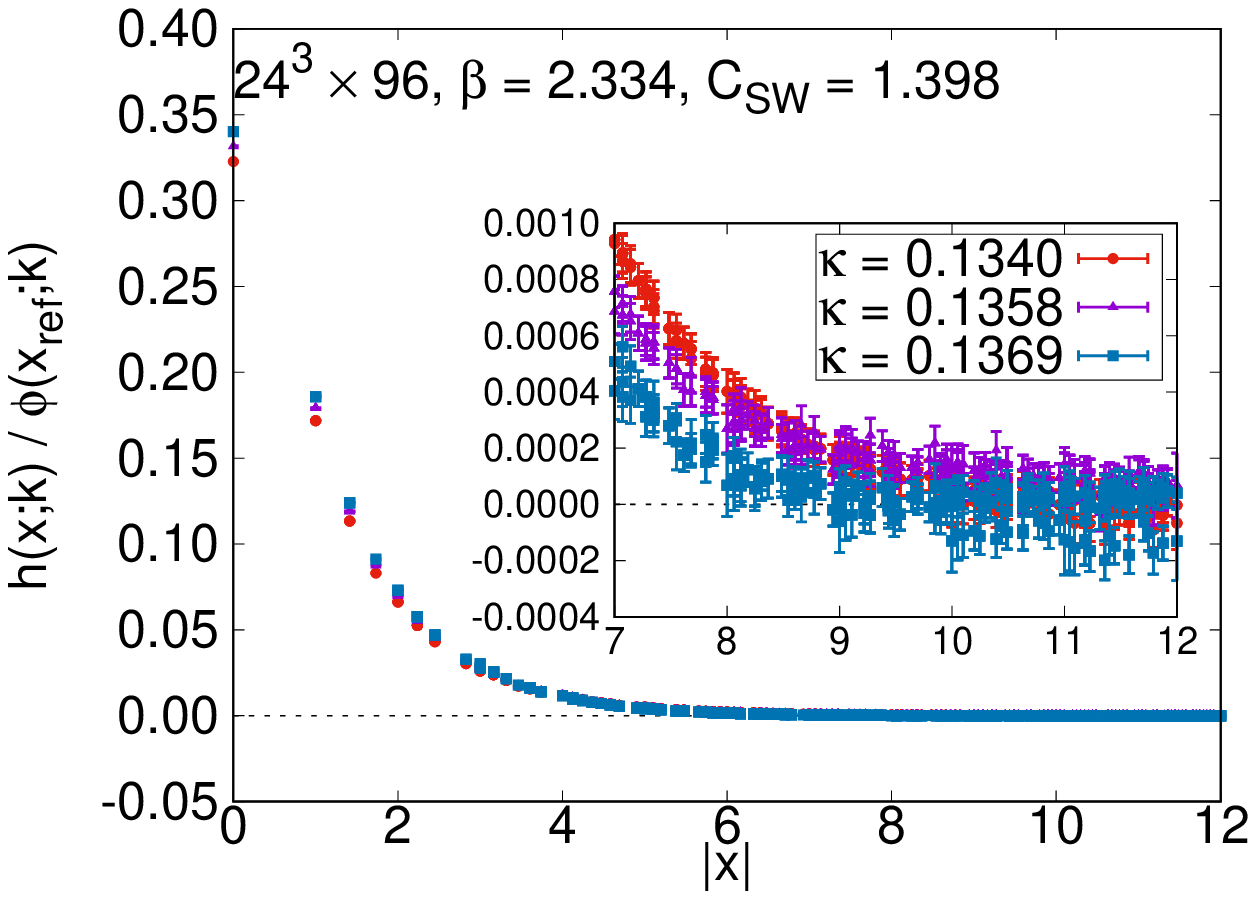}
 \includegraphics[scale=0.50]{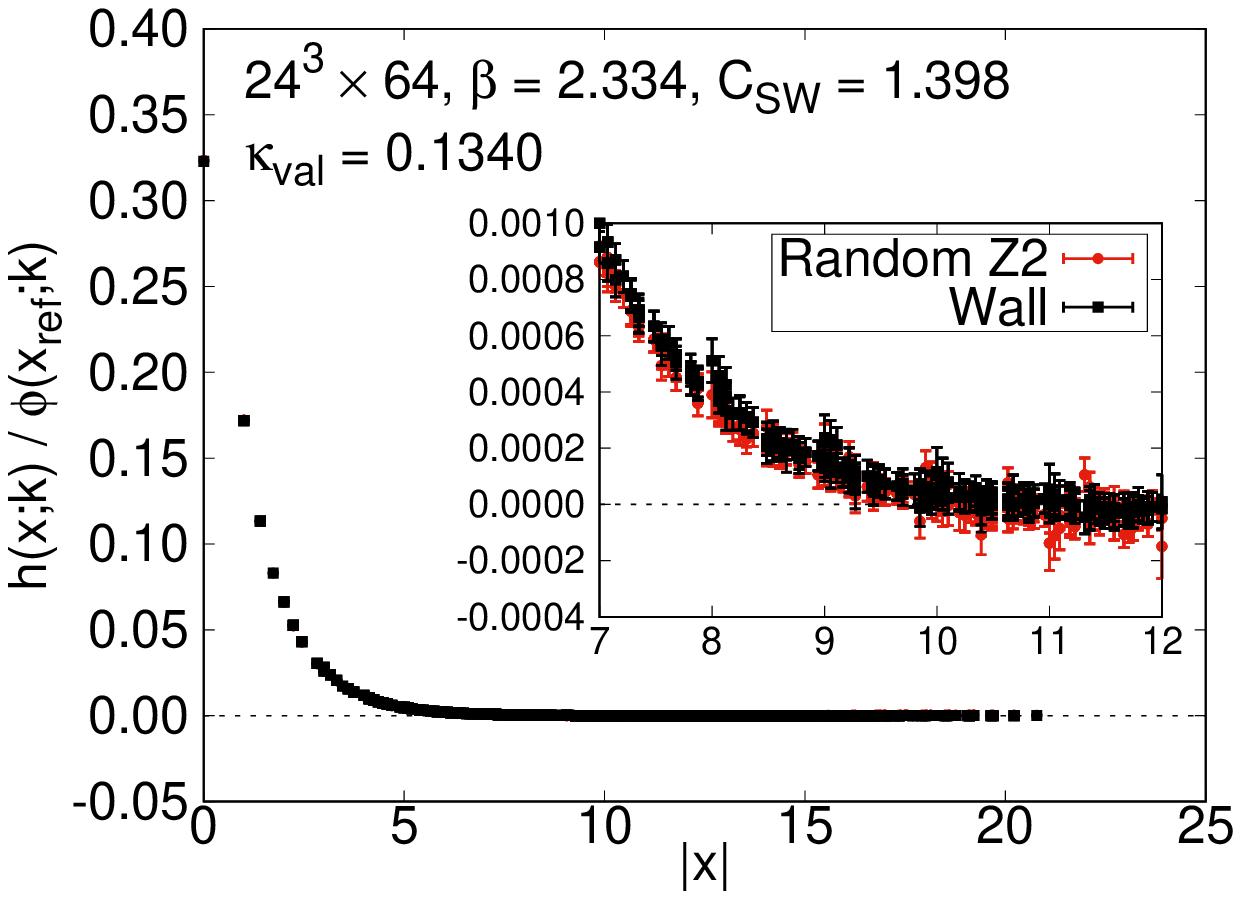}
 \caption{
  \label{fig:reduced_phi_L}
  Reduced wave functions $h({\bf x};k)$ normalized by the wave function $\phi({\bf x}_{\rm ref};k)$ at a reference point ${\bf x}_{\rm ref} = (12,7,2)$ on $24^3 \times 96$ (left panel) and on $24^3 \times 64$ (right panel).
  $k = k_t$ is employed.
 }
\end{figure*}

Figure~\ref{fig:reduced_phi_L} illustrates our results of the reduced wave function $h({\bf x};k)$ defined in Eq.~(\ref{eqn:def_hxk_lat}) with $k = k_t$.
$h({\bf x};k)$ is normalized by $\phi({\bf x};k)$ at a reference point ${\bf x}_{\rm ref}$ to eliminate the overall constant.
${\bf x}_{\rm ref} = (12,7,2)$ is chosen to minimize $l=4$ contribution to $\phi({\bf x};k)$.
The left panel of Fig.~\ref{fig:reduced_phi_L} exhibits quark mass dependence of $h({\bf x};k_s) / \phi({\bf x}_{\rm ref};k)$ on $24^3 \times 96$ with random $Z2$ sources.
Our result suggests the interaction range $R \sim 10 < L / 2$.
$h({\bf x};k) = 0$ is confirmed in $x > R$ within our statistical errors.
The sufficient condition of Eq.~(\ref{eqn:H_L_pk_lattice}) is satisfied in our quark mass region.
The right panel of Fig.~\ref{fig:reduced_phi_L} presents $h({\bf x};k_s) / \phi({\bf x}_{\rm ref};k)$ on $24^3 \times 64$ with random $Z2$ and wall sources.
Both results agree with each other.
Consistency of the results proves the source independence of $h({\bf x};k_s) / \phi({\bf x}_{\rm ref};k)$.

Satisfaction of the sufficient condition $R \sim 10 < L / 2$ allows us to evaluate $H_L(p;k)$ by Eq.~(\ref{eqn:H_L_pk_lattice}).
We use $k = k_t,k_s$ for $h({\bf x};k)$ in $H_L(p;k)$ evaluation.
We found $H_L(p;k)$ with a summation over all spatial volume is equal to that with up to $x = 10 \sim R$, suggesting our estimate of $R$ is valid.
It is noted our result still has discretization effects.
The rotational symmetry breaking appears as deviations between on-axis and off-axis data of $h({\bf x};k)$.
It causes 3\% difference in $H_L(p;k)$, which is the same order as our statistical error.
Another point is data near $x=0$ are affected more significantly by the finite lattice spacing.
Their contributions, however, are suppressed in $H_L(p;k)$ by the Jacobian $r^2$ in the integral.
$h({\bf x};k)$ at small $x$ contributes little to $H_L(p;k)$.
To remove these systematic errors, the continuum extrapolation is required.

\begin{figure}[t]
 \centering
 \includegraphics[scale=0.60]{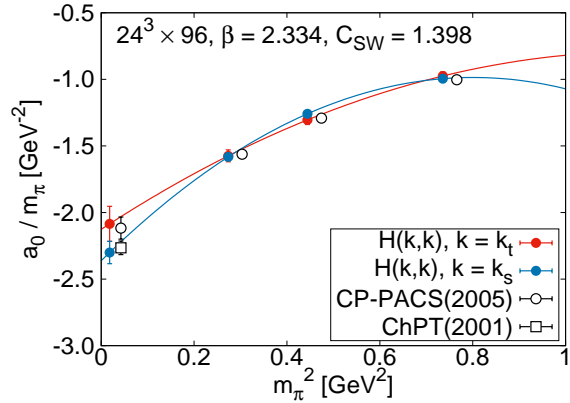}
 \caption{
  \label{fig:chiral_extrap_a_0}
  Chiral extrapolations of the scattering length over the pion mass $a_0 / m_\pi$.
  Open squares are lattice QCD results with the finite volume formula~\cite{Aoki:2005uf}.
  Open circle is a phenomenological value from ChPT~\cite{Colangelo:2001df}.
  Open symbols are slightly shifted for clarification of data.
 }
\end{figure}

At on-shell $p=k$, we determine $\tan \delta(k)$ by Eq.~(\ref{eqn:tan_delta}) and the scattering length $a_0$ by Eq.~(\ref{eqn:a_0}).
We perform a chiral extrapolation of $a_0$ to the physical point with a formula motivated by chiral perturbation theory(ChPT)~\cite{Gasser:1983yg},
\begin{eqnarray}
 a_0 / m_\pi
 = A_{a_0} + B_{a_0} m_\pi^2 + C_{a_0} m_\pi^4,
\end{eqnarray}
where $A_{a_0},B_{a_0},C_{a_0}$ are fitting parameters.
Fig.~\ref{fig:chiral_extrap_a_0} presents the chiral extrapolations of $a_0 / m_\pi$.
Our result agrees with the previous lattice QCD estimate using the finite volume formula~\cite{Aoki:2005uf} and the phenomenological value using ChPT~\cite{Colangelo:2001df}.
It validates our approach using $\phi({\bf x};k)$ inside the interaction range,
in contrast to the conventional method using $\phi({\bf x};k)$ outside the interaction range.

\begin{figure}[!tb]
 \centering
 \includegraphics[scale=0.60]{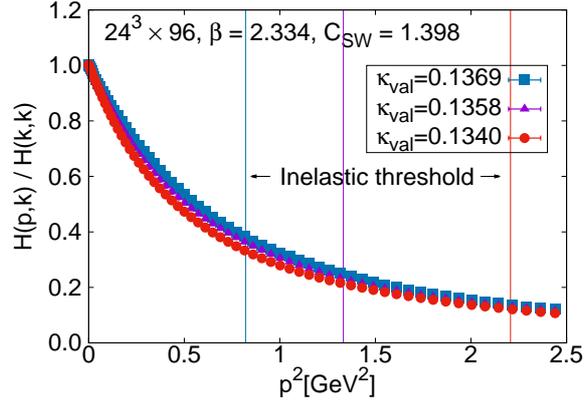}
 \caption{
  \label{fig:half_off_shell}
  Half-off-shell amplitude $H(p;k)$ normalized by its on-shell value $H(k;k)$ with $k = k_s$.
  The vertical line expresses the inelastic threshold energy.
 }
\end{figure}

Fig.~\ref{fig:half_off_shell} exhibits the half-off-shell amplitude $H(p;k)$ as a function of $p^2$, normalized by its on-shell value.
We have a clean signal of $H(p;k)$ throughout the $p^2$ range we surveyed.
The overall factor as well as the source operator dependence of $H_L(p;k)$ are canceled out in the ratio, $H_L(p;k) / H_L(k;k) = H(p;k) / H(k;k)$.
The sink smearing of the pion operator produces an extra overall factor, but it can be analytically removed~\cite{Kawai:2017goq}.
Though $H(p;k)$ is not observable directly in experiments, it is useful for effective theories and models of hadrons.
Our data of $H(p;k)$ can be an additional input to the effective theories and models, in complement to experimental data.

\begin{figure}[t]
 \centering
 \includegraphics[scale=0.60]{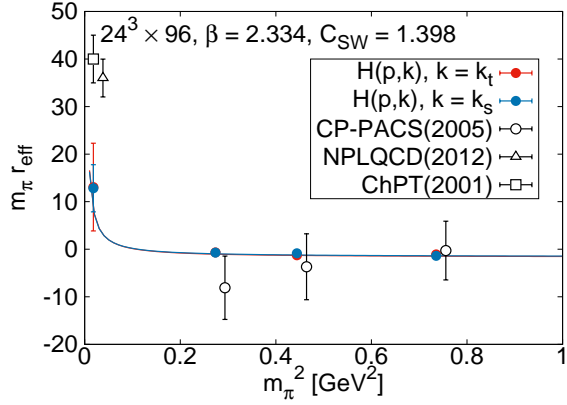}
 \caption{
  \label{fig:chiral_extrap_r_eff}
  Chiral extrapolations of the effective range $r_{\rm eff}$.
  Open squares are $N_f=0$ lattice QCD results with the finite volume formula evaluated from data in Ref.~\cite{Aoki:2005uf}.
  Open triangle is $N_f=2+1$ lattice QCD result with the finite volume formula~\cite{Beane:2011sc}.
  Open circle is a phenomenological value by ChPT~\cite{Colangelo:2001df}.
  Open symbols are slightly shifted for clarification of data.
 }
\end{figure}

The effective range $r_{\rm eff}$ is calculated with $H(p;k)$ through Eq.~(\ref{eqn:r_eff}).
Fig.~\ref{fig:chiral_extrap_r_eff} presents chiral extrapolations of $m_\pi r_{\rm eff}$.
Our results are consistent with those by the conventional finite volume formula, extracted from data in Ref.~\cite{Aoki:2005uf}.
The explicit $p^2$ dependence of $H(p;k)$ leads to more accurate results than those by the finite volume formula.
The chiral extrapolation of $r_{\rm eff}$ is performed with a formula based on ChPT~\cite{Bijnens:1997vq},
\begin{eqnarray}
 m_\pi r_{\rm eff}
 = A_{r_{\rm eff}} / m_\pi^2 + B_{r_{\rm eff}},
\end{eqnarray}
where $A_{r_{\rm eff}},B_{r_{\rm eff}}$ are fitting parameters.
At the physical point, our result clearly deviates from the phenomenological estimate~\cite{Colangelo:2001df}.
A probable possibility is the chiral extrapolation.
Data near the physical point seems to be required.
Another source is our assumptions in Eq.~(\ref{eqn:r_eff}).
The quenched approximation can also cause the deviation.
A dynamical $N_f=2+1$ lattice QCD with $m_\pi = 390$~MeV reproduces the phenomenological value~\cite{Beane:2011sc}.
More realistic $N_f=2+1$ lattice QCD data close to the physical point are needed to clarify the origin of the discrepancy.

\begin{acknowledgments}

We thank J.Carbonell and V.A.Karmnov for drawing our attention to the formulation in the momentum space.
We also thank INT at the University of Washington for the hospitality and partial financial support during an early stage of this work.
Our simulation was performed on COMA under Interdisciplinary Computational Science Program of Center for Computational Sciences, University of Tsukuba.
This work is based on Bridge++ code~\cite{Bridge}.
This work is supported in part by JSPS KAKENHI Grant Numbers 16H06002 and 18K03638.

\end{acknowledgments}

\end{document}